# Logic Learning in Hopfield Networks

Saratha Sathasivam (Corresponding author)

School of Mathematical Sciences, University of Science Malaysia,

Penang, Malaysia

E-mail: saratha@cs.usm.my

Wan Ahmad Tajuddin Wan Abdullah

Department of Physics, Universiti Malaya,

50603 Kuala Lumpur, Malaysia

E-mail: wat@um.edu.my

*The research is partly financed by an FRGS grant from the Ministry of Higher Education, Malaysia.*

**Abstract**

Synaptic weights for neurons in logic programming can be calculated either by using Hebbian learning or by Wan Abdullah's method. In other words, Hebbian learning for governing events corresponding to some respective program clauses is equivalent with learning using Wan Abdullah's method for the same respective program clauses. In this paper we will evaluate experimentally the equivalence between these two types of learning through computer simulations.

**1. Introduction**

Recurrent neural networks are essentially dynamical systems that feed back signals to themselves. Popularized by John Hopfield, these models possess a rich class of dynamics characterized by the existence of several stable states each with its own basin of attraction. The (Little-)Hopfield neural network [Little (1974), Hopfield (1982)] minimizes a Lyapunov function, also known as the energy function due to obvious similarities with a physical spin network. Thus, it is useful as a content addressable memory or an analog computer for solving combinatorial-type optimization problems because it always evolves in the direction that leads to lower network energy. This implies that if a combinatorial optimization problem can be formulated as minimizing the network energy, then the network can be used to find optimal (or suboptimal) solutions by letting the network evolve freely.

Wan Abdullah (1991,1992) and Pinkas (1991) independantly defined bi-directional mappings between propositional logic formulas and energy functions of symmetric neural networks. Both methods are applicable in finding whether the solutions obtained are models for a corresponding logic program.

Subsequently Wan Abdullah (1991, 1993) has shown on see how Hebbian learning in an environment with some underlying logical rules governing events is equivalent to hardwiring the network with these rules. In this paper, we will experimentally carry out computer simulations to support this.

This paper is organized as follows. In section 2, we give an outline of doing logic programming on a Hopfield network and in section 3, Hebbian learning of logical clauses is described. In section 4, we describe the proposed approach for comparing connection strengths obtained by Wan Abdullah's method and Hebbian learning. Section 5 contains discussions regarding the results obtained from computer simulations. Finally concluding remarks regarding this work occupy the last section.

**2. Logic Programming on a Hopfield network**

In order to keep this paper self-contained we briefly review the Hopfield model (extensive treatments can be found elsewhere [Geszti (1990), Haykin (1994)]), and how logic programming can be carried out on such architecture. The Hopfield model is a standard model for associative memory. The Hopfield dynamics is asynchronous, with each neuron updating its state deterministically. The system consists of $N$ formal neurons, each of which can be described by Ising variables $S_i(t), (i = 1, 2, ....N)$. Neurons are bipolar, $S_i \in \{-1,1\}$, obeying the dynamics $S_i \to \text{sgn}(h_i)$, where the field, $h_i = \sum_j J_{ij}^{(2)} S_j + J_i^{(1)}$, $i$ and $j$ running over all neurons $N$, $J_{ij}^{(2)}$ is the synaptic or connectio strength from neuron $j$ to neuron $i$, and $-J_i^{(1)}$ is the threshold of neuron $i$.

Restricting the connections to be symmetric and zero-diagonal, $J_{ij}^{(2)} = J_{ji}^{(2)}$, $J_{ii}^{(2)} = 0$, allows one to write a Lyapunov or energy function,

$$E = -\frac{1}{2}\sum\sum J_{ij}^{(2)} S_i S_j - \sum J_i^{(1)} S_i \qquad (1)$$

which decreases monotonically with the dynamics.

The two-connection model can be generalized to include higher order connections. This modifies the "field" into

$$h_i = .... + \sum\sum J_{ijk}^{(3)} S_j S_k + \sum J_{ij}^{(2)} S_j + J_i^{(1)} \qquad (2)$$

where "….." denotes still higher orders, and an energy function can be written as follows:

$$E = ..... - \frac{1}{3}\sum\sum\sum J_{ijk}^{(3)} S_i S_j S_k - \frac{1}{2}\sum\sum J_{ij}^{(2)} S_i S_j - \sum J_i^{(1)} S_i \qquad (3)$$

provided that $J_{ijk}^{(3)} = J_{[ijk]}^{(3)}$ for $i, j, k$ distinct, with $[...]$ denoting permutations in cyclic order, and $J_{ijk}^{(3)} = 0$ for any $i, j, k$ equal, and that similar symmetry requirements are satisfied for higher order connections. The updating rule maintains

$$S_i(t+1) = \text{sgn}[h_i(t)] \qquad (4)$$

In logic programming, a set of Horn clauses which are logic clauses of the form $A \leftarrow B_1, B_2, ..., B_n$ where the arrow may be read "if" and the commas "and", is given and the aim is to find the set(s) of interpretation (i.e., truth values for the atoms in the clauses which satisfy the clauses (which yields all the clauses true). In other words, we want to find 'models' corresponding to the given logic program.

In principle logic programming can be seen as a problem in combinatorial optimization, which may therefore be carried out on a Hopfield neural network. This is done by using the neurons to store the truth values of the atoms and writing a cost function which is minimized when all the clauses are satisfied.

As an example, consider the following logic program,

$$A \leftarrow B, C.$$
$$D \leftarrow B.$$
$$C \leftarrow .$$

whose three clauses translate respectively as $A \vee \neg B \vee \neg C$, $D \vee \neg B$ and $C$. The underlying task of the program is to look for interpretations of the atoms, in this case $A$, $B$, $C$ and $D$ which make up the model for the given logic program. This can be seen as a combinatorial optimization problem where the "inconsistency",

$$E_P = \tfrac{1}{2}(1-S_A)\tfrac{1}{2}(1+S_B)\tfrac{1}{2}(1+S_C)$$
$$+ \tfrac{1}{2}(1-S_D)\tfrac{1}{2}(1+S_B) + \tfrac{1}{2}(1-S_C) \quad (5)$$

Where $S_A$, etc. represent the truth values (*true* as 1) of $A$, etc., is chosen as the cost function to be minimized, as was done by Wan Abdullah. We can observe that the minimum value for $E_P$ is 0, and has otherwise value proportional to the number of unsatisfied clauses. The cost function (5), when programmed onto a third order neural network yields synaptic strengths as given in Table 1. We address this method of doing logic programming in neural networks as *Wan Abdullah's method*.

**3. Hebbian Learning of Logical Clauses**

The Hebbian learning rule for a two-neuron synaptic connection can be written as

$$\Delta J_{ii}^{(2)} = \lambda_2 S_i S_j \quad (6)$$

where $\lambda_2$ is a learning rate. For connections of other orders $n$, between $n$ neurons $\{S_i, S_j, ..., S_m\}$, we can generalize this to

$$\Delta J_{ii....m}^{(n)} = \lambda_n S_i S_j ....S_m \quad (7)$$

This gives the changes in synaptic strengths depending on the activities of the neurons. In an environment where selective events occur, Hebbian learning will reflect the occurrences of the events. So, if the frequency of the events is dictated by some underlying logical rule, logic should be entrenched in the synaptic weights.

Wan Abdullah (1991, 1993) has shown that Hebbian learning as above corresponds to hardwiring the neural network with synaptic strengths obtained using Wan Abdullah's method, provided that the following is true:

$$\lambda_n = \frac{1}{(n-1)!} \quad (8)$$

We do not provide a detailed analysis regarding Hebbian learning of logical clauses in this paper, but instead refer the interested reader to Wan Abdullah's papers.

**4. Comparing Connection Strengths Obtained By Hebbian Learning With Those By Wan Abdullah's Method**

In the previous section, we have elaborated how synaptic weights for neurons can be equivalently calculated either by using Hebbian learning or by Wan Abdullah's method. Theoretically, information (synaptic strengths) produced by both methods are similar. However, due to interference effects and redundancies, synaptic strengths could be different [Sathasivam (2006)], but the set of solutions for both cases should remain the same. Due to this, we cannot use direct comparison of obtained synaptic strengths. Instead, we carry out computer simulation of artificially generated logic programs and compare final states of the resulting neural networks.

To obtain the logic-programmed Hopfield network based on Wan Abdullah's method, the following algorithm is carried out:

i) Given a logic program, translate all the clauses in the logic program into basic Boolean algebraic form.

ii) Identify a neuron to each ground neuron.

iii) Initialize all connections strengths to zero.

iv) Derive a cost function that is associated with the negation of the conjuction of all the clauses, such that $\frac{1}{2}(1+S_x)$ represents the logical value of a neuron $X$, where $S_x$ is the neuron corresponding to logical atom $X$. The value of $S_x$ is defined in such a way that it carries the values of 1 if $X$ is true and -1 if $X$ is false. Negation ($X$ does not occur) is represented by $\frac{1}{2}(1-S_x)$; a conjunction logical connective is represented by multiplication whereas a disjunction connective is represented by addition.

v) Obtain the values of connection strengths by comparing the cost function with the energy.

vi) Let the neural network programmed with these connection strengths evolve until minimum energy is reached. Check whether the solution obtained is a global solution (the interpretation obtained is a model for the given logic program).

We run the relaxation for 1000 trials and 100 combinations of neurons so as to reduce statistical error. The selected tolerance value is 0.001. All these values are obtained by try and error technique, where we tried several values as tolerance values, and selected the value which gives better performance than other values. To compare the information obtain in the synaptic strength, we make comparison between the stable states (states in which no neuron changes its value anymore) obtained by Wan Abdullah's method with stable states obtained by Hebbian learning. The way we calculated the percentage of solutions reaching the global solutions is by comparing the energy for the stable states obtained by using Hebbian learning and Wan Abdullah's method. If the corresponding energy for both learning is same, then we conclude that the stable states for both learning are the same. This indicates, the model (set of interpretations) obtained for both learning are similar. In all this, we assume that the global solutions for both networks are the same due to both methods considering the same knowledge base (clauses).

**5. Results and Discussion**

Figures 1 - 6 illustrate the graphs for global minima ratio (ratio= (Number of global solutions)/ (Number of solutions=number of runs)) and Hamming distances from computer simulation that we have carried out. From the graphs obtained, we observed that the ratio of global solutions is consistently 1 for all the cases, although we increased the network complexity by increasing the number of neurons (NN) and number of literals per clause (NC1, NC2, NC3). Due to we are getting similar results for all the trials, to avoid graphs overlapping, we only presented the result obtained for the number of neurons (NN) = 40. Besides that, error bar for some of the cases could not be plotted because the size of the point is bigger than the error bar. This indicates that the statistical error for the corresponding point is so small. So, we couldn't plot the error bar.

Most of the neurons which are not involved in the clauses generated will be in the global states. The random generated program clause relaxed to the final states, which seem also to be stable states, in less than five runs. Furthermore, the network never gets stuck in any suboptimal solutions. This indicates good solutions (global states) can be found in linear time or less with less complexity.

Since all the solutions we obtained are global solution, so the distance between the stable states and the attractors are zero. Supporting this, we obtained zero values for Hamming distance. This indicates the stable states for both learning are the same. Therefore they are no different in the energy value. So, models for both learning are proved to be similar. Although the way of calculating synaptic weights are different, since the calculations revolve around the same knowledge base (clauses), the set of interpretations will be similar. This implies that, Hebbian learning could extract the underlying logical rules in a given set of events and provide good solutions as well as Wan Abdullah's method. The computer simulation results support this hypothesis.

**6. Conclusion**

In this paper, we had evaluated experimentally the logical equivalent between these two types of learning (Wan Abdullah's method and Hebbian learning) for the same respective clauses (same underlying logical rules) using computer simulation. The results support Wan Abdullah's earlier proposed theory.

**References**

Geszti, T. (1990). *Physical Models of Neural Networks*. Singapore: World Scientific Publication.

Haykin, S. (1994). *Neural Network: A Comprehensive Foundation*. New York: Macmillan.

Hopfield, J. J. (1982). Neural Networks and Physical Systems with Emergent Collective Computational abilities. *Proc. Natl. Acad. Sci. USA,* 79, 2554-2558.

Little, W. A. (1974). The existence of persistent states in the brain. *Math. Biosci.*, 19, 101-120.

Pinkas, G. (1991). Energy minimization and the satisfiability of propositional calculus. *Neural Computation,* 3, 282-291.

Sathasivam, S. (2006). Logic Mining in Neural Networks. PhD Thesis. University of Malaya, Malaysia.

Wan Abdullah, W. A. T. (1991). Neural Network logic. In O. Benhar et al. (Eds.), *Neural Networks: From Biology to High Energy Physics*. Pisa: ETS Editrice. pp. 135-142.

Wan Abdullah, W. A. T. (1992). Logic programming on a neural network. *Int. J. Intelligent Sys*., 7, 513-519.

Wan Abdullah, W. A. T. (1993). The logic of neural networks. *Phys. Lett.*,176A, 202-206.

Table 1: Synaptic strengths for $A \leftarrow B, C.$ & $D \leftarrow B$ & $C \leftarrow$ using Wan Abdullah's method

| Synaptic Strengths | Clause | | | Total |
|---|---|---|---|---|
| | $A \leftarrow B, C.$ | $D \leftarrow B.$ | $C \leftarrow .$ | |
| $J^{(3)}_{[ABC]}$ | 1/16 | 0 | 0 | 1/16 |
| $J^{(3)}_{[ABD]}$ | 0 | 0 | 0 | 0 |
| $J^{(3)}_{[ACD]}$ | 0 | 0 | 0 | 0 |
| $J^{(3)}_{[BCD]}$ | 0 | 0 | 0 | 0 |
| $J^{(2)}_{[AB]}$ | 1/8 | 0 | 0 | 1/8 |
| $J^{(2)}_{[AC]}$ | 1/8 | 0 | 0 | 1/8 |
| $J^{(2)}_{[AD]}$ | 0 | 0 | 0 | 0 |
| $J^{(2)}_{[BC]}$ | -1/8 | 0 | 0 | -1/8 |
| $J^{(2)}_{[BD]}$ | 0 | 1/4 | 0 | ¼ |
| $J^{(2)}_{[CD]}$ | 0 | 0 | 0 | 0 |
| $J^{(1)}_{[A]}$ | 1/8 | 0 | 0 | 1/8 |
| $J^{(1)}_{[B]}$ | -1/8 | -¼ | 0 | -3/8 |
| $J^{(1)}_{[C]}$ | -1/8 | 0 | 1/2 | 3/8 |
| $J^{(1)}_{[D]}$ | 0 | 1/4 | 0 | ¼ |

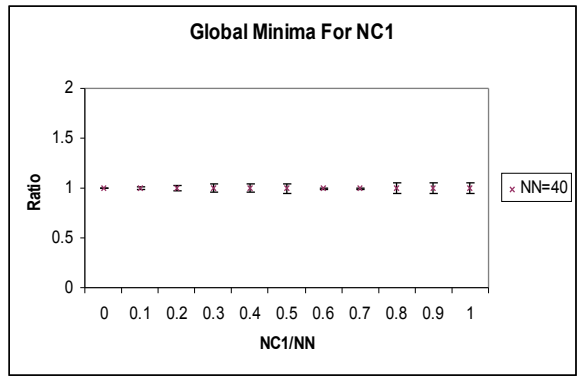

Figure 1: Global Minima Ratio for NC1

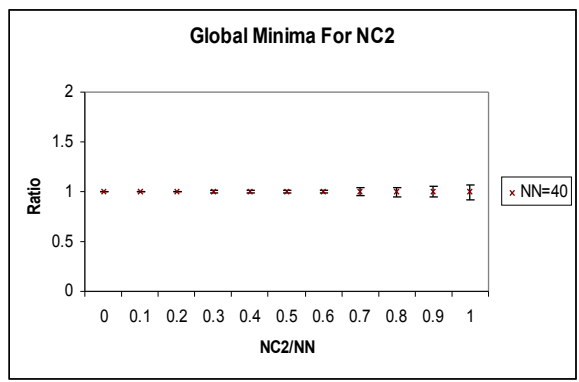

Figure 2: Global Minima Ratio for NC2

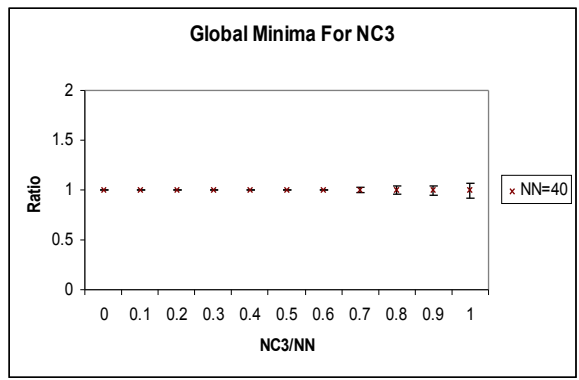

Figure 3: Global Minima Ratio for NC3

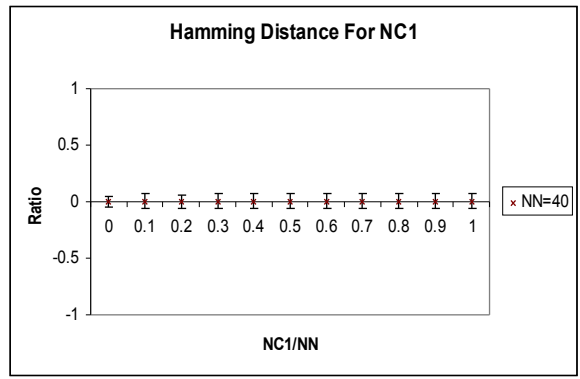

Figure 4: Hamming Distance for NC1

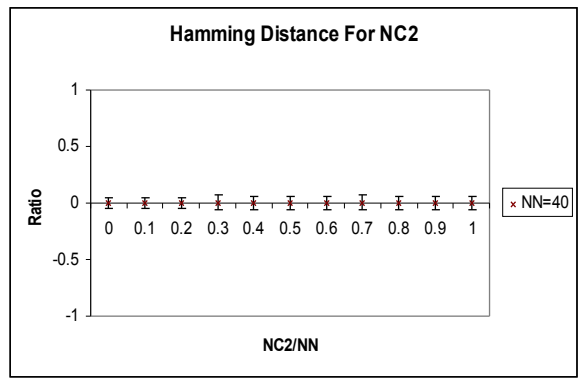

Figure 5: Hamming Distance for NC2

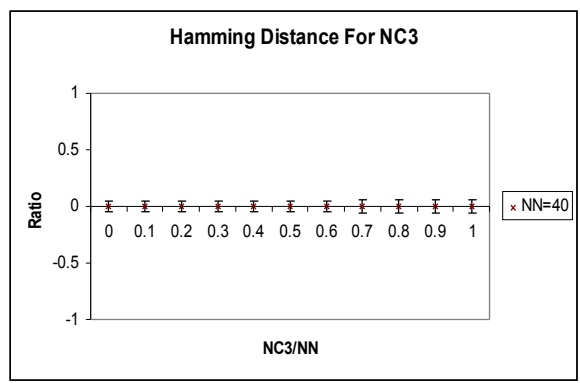

Figure 6: Hamming Distance for NC3